\documentclass[aip,reprint,author-year,nofootinbib,floatfix]{revtex4-2}

\usepackage{physjour_aas_macros}

\usepackage{graphicx}
\usepackage{amsmath,amssymb}             
\usepackage{stmaryrd}					
\usepackage[utf8]{inputenc}
\usepackage[T1]{fontenc}
\usepackage[dvipsnames]{xcolor}
\usepackage{color}
\usepackage[raggedright]{titlesec}
\usepackage[normalem]{ulem}
\definecolor{darkblue}{rgb}{0.0,0.0,0.4}
\definecolor{darkred}{rgb}{0.7,0.0,0.0}
\definecolor{darkgreen}{rgb}{0.0,0.5,0.0}
\definecolor{C0}{HTML}{1f77b4}
\definecolor{C1}{HTML}{ff7f0e}
\definecolor{C2}{HTML}{2ca02c}
\definecolor{C3}{HTML}{d62728}
\definecolor{C4}{HTML}{9467bd}
\definecolor{C5}{HTML}{8c564b}
\definecolor{C6}{HTML}{e377c2}
\definecolor{C7}{HTML}{7f7f7f}
\definecolor{C8}{HTML}{bcbd22}
\definecolor{C9}{HTML}{17becf}

\usepackage{hyperref}
\hypersetup{colorlinks,breaklinks,
            linkcolor=darkblue,urlcolor=darkblue,
            anchorcolor=darkblue,citecolor=RoyalBlue}

\usepackage[mathscr]{euscript}
\usepackage{upgreek}
\DeclareFontFamily{OT1}{pzc}{}
\DeclareFontShape{OT1}{pzc}{m}{it}{<-> s * [1.10] pzcmi7t}{}
\DeclareMathAlphabet{\mathpzc}{OT1}{pzc}{m}{it}

\titleformat{\section}{\selectfont \normalfont\raggedright\sffamily\small\bfseries\uppercase}{\thesection.}{1em}{}{}
\titleformat{\subsection}{\selectfont \normalfont\raggedright\sffamily\small\bfseries}{\thesubsection.}{1em}{}{}
\titleformat{\subsubsection}{\selectfont \normalfont\sffamily\small\bfseries}{\thesubsection.\thesubsubsection}{1em}{}{}
\titleformat{\paragraph}[runin]{\selectfont \normalfont\sffamily\small\bfseries}{\thesubsection.\thesubsubsection.\theparagraph}{1em}{}[.]

\usepackage{tikz}
\usetikzlibrary{shapes,arrows,shadows}
\usepackage[export]{adjustbox}

\newcommand{\deriv}[2]{\dfrac{\mathrm{d} #1}{\mathrm{d} #2}}

\newcommand{\drm}{\mathrm{d}}
\newcommand{\erm}{\mathrm{e}}

\newcommand{\p}{\mathpzc{P}}

\DeclareMathOperator{\tr}{tr}

\newcommand{\XiG}{\boldsymbol{\Xi}_\mathrm{G}}
\newcommand{\XiGbeta}{\boldsymbol{\Xi}_\mathrm{G,\beta}}
\newcommand{\XiGbetabeta}{\boldsymbol{\Xi}_\mathrm{G,\beta\beta}}
\newcommand{\Fxi}{\textbf{F}^{\xi}}
\newcommand{\Ff}{\textbf{F}^{f}}
\newcommand{\Mij}{\textbf{M}_{ij}}
\newcommand{\boF}{\textbf{F}}
\newcommand{\boK}{\textbf{K}}
\newcommand{\boQ}{\textbf{Q}}
\newcommand{\boR}{\textbf{R}}
\newcommand{\boS}{\textbf{S}}
\newcommand{\boSig}{\boldsymbol{\Sigma}}
\newcommand{\boSighat}{\boldsymbol{\hat{\Sigma}}}
\newcommand{\bof}{f}
\newcommand{\bog}{g}
\newcommand{\boh}{h}
\newcommand{\boj}{j}
\newcommand{\boy}{y}
\newcommand{\boxi}{\xi}
\newcommand{\botheta}{\boldsymbol{\uptheta}}

\newcommand{\comment}[1]{}

\definecolor{LBAcolor}{HTML}{51abe8}
\definecolor{DAcolor}{HTML}{ffb43f}
\definecolor{SBIcolor}{HTML}{5ed45f}

\begin{document}


\title{On the accuracy and precision of correlation functions and field-level inference in cosmology}



\author{Florent~Leclercq}
\email{florent.leclercq@polytechnique.org}
\homepage{https://www.florent-leclercq.eu/}
\affiliation{Imperial Centre for Inference and Cosmology (ICIC) \& Astrophysics Group, Imperial College London, Blackett Laboratory, Prince Consort Road, London SW7 2AZ, United Kingdom}

\author{Alan~Heavens}
\affiliation{Imperial Centre for Inference and Cosmology (ICIC) \& Astrophysics Group, Imperial College London, Blackett Laboratory, Prince Consort Road, London SW7 2AZ, United Kingdom}


\date{\today}

\begin{abstract}
We present a comparative study of the accuracy and precision of correlation function methods and full-field inference in cosmological data analysis. To do so, we examine a Bayesian hierarchical model that predicts log-normal fields and their two-point correlation function. Although a simplified analytic model, the log-normal model produces fields that share many of the essential characteristics of the present-day non-Gaussian cosmological density fields. We use three different statistical techniques: (\textit{i}) a standard likelihood-based analysis of the two-point correlation function; (\textit{ii}) a likelihood-free (simulation-based) analysis of the two-point correlation function; (\textit{iii}) a field-level analysis, made possible by the more sophisticated data assimilation technique. We find that (a) standard assumptions made to write down a likelihood for correlation functions can cause significant biases, a problem that is alleviated with simulation-based inference; and (b) analysing the entire field offers considerable advantages over correlation functions, through higher accuracy, higher precision, or both. The gains depend on the degree of non-Gaussianity, but in all cases, including for weak non-Gaussianity, the advantage of analysing the full field is substantial.
\end{abstract}


\maketitle



\section{Introduction}

As cosmological surveys increase in size, the statistical errors decrease and the possibility for new discoveries increases.  This increase in precision needs to be accompanied by equally high accuracy in order to avoid the wrong inferences being drawn.  With the imminent arrival of the Euclid satellite \citep{Euclid} and the Legacy Survey of Space and Time \citep{LSSTScienceCollaboration2012}, the time is right to assess whether the standard methods and more advanced tools for statistical analysis of late-time, non-Gaussian, cosmological fields are likely to be adequate. In this letter, we compare methods by analysing an analytic model, namely the log-normal (LN) model, that has many of the salient features of the late-time fields such as the galaxy density field or the weak lensing cosmic shear field.

We focus on three analysis methods: (\textit{i}) the most common standard technique, i.e. a likelihood-based analysis (LBA) of the two-point correlation function (2PCF), assuming a Gaussian distribution with fixed covariance matrix; (\textit{ii}) the relatively new technique (to cosmology) of simulation-based inference (SBI), also known as likelihood-free inference, based on the 2PCF; (\textit{iii}) and the computationally expensive but powerful data assimilation (DA) technique, which allows field-level inference without any compression of the data. We investigate fields of varying levels of non-Gaussianity, and find that field-level inference offers advantages in accuracy and precision over the 2PCF that can be extremely large. For significantly non-Gaussian fields, the precision of LBA and SBI are similar, but SBI is typically more accurate (less ``biased''). DA, which uses all the field values and captures all of the information, outperforms both and provides highly accurate and precise results. We also study weakly non-Gaussian fields, where one might expect the precision of 2PCF analyses to equal that of field-level inference, as the common lore is that the statistical properties in the limit of vanishing non-Gaussianity are completely captured by the 2PCF. This turns out not to be the case, and, although we find comparable statistical errors in the shape parameter of field correlations, DA gives much more precise measurements of the degree of non-Gaussianity.

\section{Model}
\label{sec:Model}

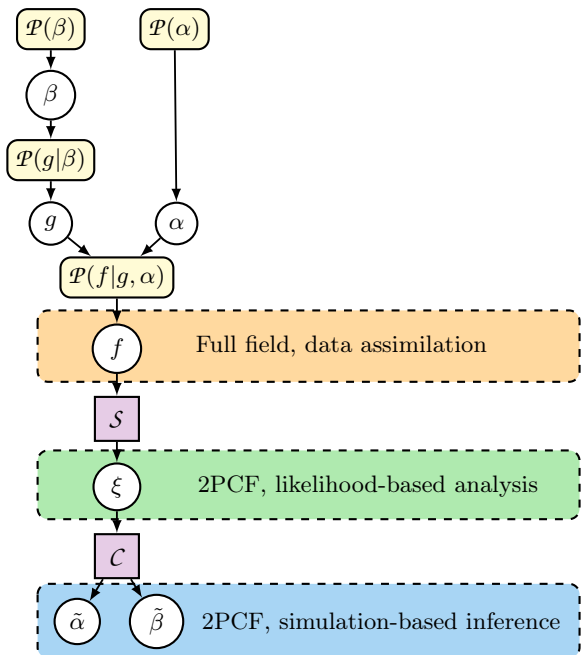
\begin{figure}

\begin{center}
\begin{tikzpicture}
	\pgfdeclarelayer{background}
	\pgfdeclarelayer{foreground}
	\pgfsetlayers{background,main,foreground}

	\tikzstyle{probability}=[draw, thick, text centered, rounded corners, minimum height=1em, minimum width=1em, fill=yellow!20]
	\tikzstyle{deterministic}=[draw, thick, text centered, minimum height=1.8em, minimum width=1.8em, fill=violet!20]
	\tikzstyle{variabl}=[draw, thick, text centered, circle, minimum height=1em, minimum width=1em, fill=white]
	\tikzstyle{FLI} = [draw, thick, rectangle, rounded corners, dashed, fill=DAcolor!50]
	\tikzstyle{LBA} = [draw, thick, rectangle, rounded corners, dashed, fill=SBIcolor!50]
	\tikzstyle{SBI} = [draw, thick, rectangle, rounded corners, dashed, fill=LBAcolor!50]

	\def\blockdist{0.7}
	\def\modeldist{2.0}

    \node (betaproba) [probability]
    {$\p(\beta)$};
    \path (betaproba.south)+(0,-17pt) node (beta) [variabl]
    {$\beta$};
    \path (beta.south)+(0,-15pt) node (fgproba) [probability]
    {$\p(g|\beta)$};
    \path (fgproba.south)+(0,-16pt) node (fg) [variabl]
    {$g$};
    \path (betaproba.west)+(60pt,0) node (alphaproba) [probability]
    {$\p(\alpha)$};
    \path (fg.west)+(56pt,0) node (alpha) [variabl]
    {$\alpha$};
    \path (fg.south)+(25pt,-12pt) node (flnproba) [probability]
    {$\p(f|g,\alpha)$};
	\path [FLI] (flnproba.south)+(-30pt,-4.5pt) rectangle (200pt,-133.5pt)
    {};
	\node at (110pt,-120pt) {Full field, data assimilation};
    \path (flnproba.south)+(0,-19pt) node (fln) [variabl]
    {$f$};
    \path (fln.south)+(0,-17pt) node (S) [deterministic]
    {$\mathpzc{S}$};
    \path [LBA] (flnproba.south)+(-30pt,-57.5pt) rectangle (200pt,-185.5pt)
    {};
	\node at (120pt,-173pt) {2PCF, likelihood-based analysis};
    \path (S.south)+(0,-17pt) node (xi) [variabl]
    {$\xi$};
    \path (xi.south)+(0,-17pt) node (C) [deterministic]
    {$\mathpzc{C}$};
    \path [SBI] (flnproba.south)+(-30pt,-108pt) rectangle (200pt, -238pt)
    {};
	\node at (125pt,-225pt) {2PCF, simulation-based inference};
    \path (C.south)+(-15pt,-16pt) node (alphatilde) [variabl]
    {$\tilde{\alpha}$};
    \path (C.south)+(15pt,-16pt) node (betatilde) [variabl]
    {$\tilde{\beta}$};

	\path [draw, line width=0.7pt, arrows={-latex}] (betaproba) -- (beta);
	\path [draw, line width=0.7pt, arrows={-latex}] (beta) -- (fgproba);
	\path [draw, line width=0.7pt, arrows={-latex}] (fgproba) -- (fg);
	\path [draw, line width=0.7pt, arrows={-latex}] (alphaproba) -- (alpha);
	\path [draw, line width=0.7pt, arrows={-latex}] (fg) -- (flnproba);
	\path [draw, line width=0.7pt, arrows={-latex}] (alpha) -- (flnproba);
	\path [draw, line width=0.7pt, arrows={-latex}] (flnproba) -- (fln);
	\path [draw, line width=0.7pt, arrows={-latex}] (fln) -- (S);
	\path [draw, line width=0.7pt, arrows={-latex}] (S) -- (xi);
	\path [draw, line width=0.7pt, arrows={-latex}] (xi) -- (C);
	\path [draw, line width=0.7pt, arrows={-latex}] (C) -- (alphatilde);
	\path [draw, line width=0.7pt, arrows={-latex}] (C) -- (betatilde);

\end{tikzpicture}
\end{center}

    \vspace*{-10pt}
    \caption{Graphical representation of the Bayesian hierarchical model used in this work. The rounded yellow boxes represent probability distributions and the purple squares represent deterministic functions. Three different statistical methods, as indicated by dashed rectangles, are considered.}
    \label{fig:lognormal_bhm_graph}
    \vspace*{-4pt}
\end{figure}

We consider discrete two-dimensional log-normal random fields \citep{Coles1991} of size $N_\mathrm{pix}$. Such fields can be seen as variables of a Bayesian hierarchical model (BHM) represented in figure~\ref{fig:lognormal_bhm_graph}. Specifically, the starting point is a Gaussian random field (GRF) $\bog$ \citep[see][]{Peacock1985,Bardeen1986} of size $N_\mathrm{pix}$, with zero mean and a covariance matrix $\XiG$. We further assume that the 2PCF $\xi_\mathrm{G}(r)$, giving the coefficients of $\XiG$, is parametrised by a single variable $\beta$, such that 
\begin{equation}
\p(\bog|\beta) = \mathpzc{G} \left[ \boldsymbol{0}, \XiG \right] \quad \mathrm{with} \quad \xi_\mathrm{G}(r) = \exp\left(-\frac{1}{4} \frac{r^2}{\beta^2}\right),
\label{eq:p_f_G}
\end{equation}
where $r$ is the separation between two grid points, in units of pixels. The GRF $\bog$ is made log-normal via the following transform:
\begin{equation}
\mathrm{LN}(\bog,\alpha) \equiv \frac{1}{\alpha} \left[\exp\left(\alpha \bog - \frac{1}{2} \alpha^2\right) -1 \right],
\label{eq:LN}
\end{equation}
which involves one free parameter $\alpha$. The final field $\bof$ is assumed to be a noisy realisation, the expectation value being given by equation \eqref{eq:LN}. For simplicity, we assume a Gaussian noise model, uniform in real space, with standard deviation $\sigma$. Therefore,
\begin{equation}
\p(\bof|\bog,\alpha) = \mathpzc{G} \left[ \mathrm{LN}(\bog,\alpha), \mathrm{diag}(\sigma^2) \right].
\label{eq:p_f_LN}
\end{equation}
We note that in equation \eqref{eq:LN}, the term $-\frac{1}{2}\alpha^2$ in the exponential ensures that the mean of $\mathrm{LN}(\bog,\alpha)$ is zero,\footnote{Normally this term is multiplied by the variance of the Gaussian field, which is unity by construction.} and the prefactor $\frac{1}{\alpha}$ ensures that the signal to noise ratio in $\bof$ is independent of $\alpha$, in the limit $\alpha \rightarrow 0$.

Remarkably, in the absence of noise, LN fields have analytic correlation functions at all orders \citep{Coles1991}. In particular, their 2PCF is given by
\begin{equation}
\xi_\mathrm{LN}(r) \equiv \frac{1}{\alpha^2} \left\{ \exp\left[\alpha^2 \exp\left(-\frac{1}{4} \frac{r^2}{\beta^2}\right) \right] - 1 \right\},
\label{eq:xi_LN}
\end{equation}
whose gradients with respect to $\alpha$ and $\beta$ are readily obtained (see appendix \ref{apx:Gradients of the log-normal 2PCF}).

The equations above can be interpreted as a simple model for observations of the matter distribution in the Universe. As can be observed in equation \eqref{eq:p_f_G}, $\beta$ captures the strength of correlations in $\bog$, and can therefore be seen as a proxy for the initial matter power spectrum. From equation \eqref{eq:LN}, it appears that $\alpha$ characterises the strength of non-Gaussianities in the final field, and therefore it can been seen as a proxy for the non-linear evolution of the matter field. Note that the signal is arbitrarily close to the original Gaussian random field $\bog$ at early times, i.e. in the limit $\alpha \rightarrow 0$. All other physical and observational processes are modelled stochastically by the additive Gaussian noise. 

Throughout this paper, we adopt $N_\mathrm{pix}=20^2$, $\sigma = 0.01$, and uniform (hyper-)priors $\p(\alpha)$ and $\p(\beta)$ on the intervals $[0,2]$ and $[0.2,0.8]$, respectively. We aim at inferring $\alpha$ and $\beta$ from observations of one or several realisations of $\bof$, using different techniques.

\section{Likelihood-based analysis of the 2PCF}
\label{sec:LBA}

The first technique we consider is usual in cosmology, and consists in a LBA of the estimated 2PCF $\boxi^\mathrm{obs}$. Given a realisation $\bof^\mathrm{obs}$ of the model, an estimator of the 2PCF is constructed by taking the outer product of $\bof^\mathrm{obs}$ and itself, and then averaging these two-point correlations in different $r$-bins. This is a deterministic process, denoted $\mathpzc{S}$, that compresses any field $\bof$ into a new data vector $\boxi$ of size $N_\xi$, where $N_\xi$ is the number of $r$-bins. Formally, this can be seen as adding a layer at the bottom of the BHM, shown in green in figure~\ref{fig:lognormal_bhm_graph}.

Without any prior knowledge of the sampling distribution $\p(\boxi|\alpha,\beta)$, we make the common assumption that it is a multivariate Gaussian distribution,
\begin{equation}
\p(\boxi|\alpha,\beta,\boSig) = \mathpzc{G} \left[ \boxi_\mathrm{LN}(\alpha,\beta),\boSig \right],
\label{eq:LBA_Gaussian_assumption}
\end{equation}
where the expectation value is given by the analytic form of equation \eqref{eq:xi_LN} and the covariance matrix is denoted $\boSig$. With the LN model, it would be possible to compute $\boSig$ analytically using the expression of the four-point function. However, in this paper we adopt the more general scenario where $\boSig$ is unknown and can only be evaluated through the use of simulations. We further make the common assumption that $\boSig$ does not depend on $\alpha$ and $\beta$. Given $N_\mathrm{sims}$ realisations of $\boxi$, denoted $\boxi_i$ ($1 \leq i \leq  N_\mathrm{sims}$), $\boSig$ is evaluated using the usual unbiased estimator
\begin{align}
& \boSighat \equiv \frac{1}{N_\mathrm{sims}-1} \sum_{i=1}^{N_\mathrm{sims}} \left( \boxi_i -\bar{\boxi} \right)\left( \boxi_i -\bar{\boxi} \right)^\intercal, \label{eq:covariance_estimator}\\
& \mathrm{where} \;\; \bar{\boxi} \equiv \frac{1}{N_\mathrm{sims}}\sum_{i=1}^{N_\mathrm{sims}} \boxi_i. \nonumber
\end{align}
When estimating parameters from Gaussian-distributed data in situations where the covariance matrix is unknown and estimated from simulations, the Gaussian likelihood should be replaced by a multivariate $t$-distribution \citep{SellentinHeavens2016}. Thereby, the final likelihood that we use satisfies
\begin{equation}
\p(\boxi|\alpha,\beta,\boSighat) \propto \left[ 1 + \frac{\left( \boxi - \boxi_\mathrm{LN} \right)^\intercal \boSighat^{-1} \left( \boxi -\boxi_\mathrm{LN} \right)}{N_\mathrm{sims}-1} \right]^{-\frac{N_\mathrm{sims}}{2}} .
\label{eq:LBA_posterior}
\end{equation}
With the estimated 2PCF $\boxi^\mathrm{obs}$ and uniform priors, the LBA posterior is thus $\p_\mathrm{LBA}(\alpha,\beta|\boxi^\mathrm{obs}) \propto \p(\boxi^\mathrm{obs}|\alpha,\beta,\boSighat)$.

\section{Simulation-based inference using the 2PCF}
\label{sec:LFI}

The second technique we consider is SBI, that is to say, a likelihood-free analysis of the estimated 2PCF $\boxi^\mathrm{obs}$. Since we aim at comparing different inference techniques, the model is the same as in sections \ref{sec:Model} and \ref{sec:LBA} for the generation of LN fields and the estimation of $\boxi$. Only the statistical assumptions differ.

SBI is known to be difficult when the size of the data vector is large. To solve this issue, we add a compression step, denoted by $\mathpzc{C}$. Resulting compressed variables correspond to an additional layer at the bottom of the BHM, represented in blue in figure~\ref{fig:lognormal_bhm_graph}. In this work, we use a compression using the score function, a generalisation of the MOPED algorithm \citep{Heavens2000,AlsingWandelt2018}. Specifically, to compress the data, we assume that the sampling distribution of $\boxi$ is Gaussian-distributed, i.e.
\begin{equation}
\mathcal{L} \equiv \ln \p(\boxi|\alpha,\beta,\boSig) = -\frac{1}{2} (\boxi - \boxi_\mathrm{LN})^\intercal \boSig^{-1} (\boxi - \boxi_\mathrm{LN}) - \frac{1}{2} \ln \vert \boSig \vert .
\label{eq:log_likelihood_compression}
\end{equation}
As we further assume that $\boSig$ is independent of parameters ($\nabla \boSig = 0$), the Fisher information matrix takes a simple form:
\begin{equation}
\boF \equiv -\mathrm{E}\left[ \nabla\nabla \mathcal{L} \right] = \nabla \boxi_\mathrm{LN}^\intercal \boSig^{-1} \nabla \boxi_\mathrm{LN}.
\label{eq:fisher_definition_gaussian}
\end{equation}
Using a fiducial point $\botheta_* = (\alpha_*, \beta_*)$ in parameter space, a quasi maximum-likelihood estimator for the parameters is $\tilde{\botheta} \equiv \botheta_* + \boF^{-1}_* \nabla \mathcal{L}_*$, where the inverse of the Fisher matrix and the gradient of the log-likelihood are evaluated at the fiducial point. In our case, the expression is therefore \citep{{AlsingWandelt2018}}
\begin{equation}
\tilde{\botheta} = \botheta_* + \boF^{-1}_* \left[ \nabla \boxi_\mathrm{LN*}^\intercal \boSighat^{-1} (\boxi - \boxi_\mathrm{LN*}) \right],
\label{eq:compression_mle}
\end{equation}
with
\begin{equation}
\boF_* = \nabla \boxi_\mathrm{LN*}^\intercal \boSighat^{-1} \nabla \boxi_\mathrm{LN*}.
\label{eq:compression_fisher}
\end{equation}
The quasi maximum-likelihood estimator $\tilde{\botheta}$ compresses a data vector of arbitrary size to $p$ summaries, where $p$ is the number of target parameters of the problem (two in our case). It is an optimal compression, in the sense that it saturates the lower bound of the Fisher information inequality. Compressing simulated data $\boxi$ via equations \eqref{eq:compression_mle} and \eqref{eq:compression_fisher} requires only the estimated covariance matrix $\boSighat$ already used for the LBA (equation \eqref{eq:covariance_estimator}), and the gradients of the expectation value $\nabla \boxi_\mathrm{LN}^\intercal$, for which we use the analytic expressions (equations \eqref{eq:dxi_LN_dalpha} and \eqref{eq:dxi_LN_dbeta}). It yields two numbers that we note $\tilde{\alpha}$ and $\tilde{\beta}$. Applying the same compressor $\mathpzc{C}$ to $\boxi^\mathrm{obs}$ yields $\tilde{\alpha}^\mathrm{obs}$ and $\tilde{\beta}^\mathrm{obs}$. We note that, in this section, the Gaussian assumption of equation \eqref{eq:log_likelihood_compression} is used only for compression and not for subsequent inference. In SBI, compression can be performed under an approximate likelihood without introducing biases in the result. 

By definition, SBI shall only rely on forward evaluations of the data model. In the last few years, sophisticated algorithms have been developed to drastically reduce the number of simulations needed and/or to scale to high dimension \citep{Leclercq2018BOLFI,Alsing2018,Alsing2019,Leclercq2019SELFI}. Given that we are only interested in a two-parameter problem and that our simulations are computationally very cheap, in this paper we rely on the simplest solution for SBI, namely likelihood-free rejection sampling (sometimes also known as Approximate Bayesian Computation). To do so, we replace the parametric assumption for the likelihood $\p(\boxi|\alpha,\beta)$ (equation \eqref{eq:LBA_posterior}) by a measurement of the discrepancy between simulated $\{\tilde{\alpha},\tilde{\beta}\}$ and observed $\{\tilde{\alpha}^\mathrm{obs},\tilde{\beta}^\mathrm{obs}\}$ compressed data. We choose the Euclidean distance,
\begin{equation}
\Delta(\tilde{\alpha},\tilde{\beta}) \equiv \sqrt{(\tilde{\alpha}-\tilde{\alpha}^\mathrm{obs})^2 + (\tilde{\beta}-\tilde{\beta}^\mathrm{obs})^2}.
\end{equation}
Given a small threshold $\varepsilon$, the algorithm for likelihood-free rejection sampling is then to iterate many times the following procedure: draw $(\alpha,\beta)$ from a proposal distribution (in our case, the uniform priors), simulate $\tilde{\alpha}$ and $\tilde{\beta}$ using the full BHM represented in figure~\ref{fig:lognormal_bhm_graph}, compute the discrepancy $\Delta(\tilde{\alpha},\tilde{\beta})$, then accept $(\alpha,\beta)$ as a sample of the approximate posterior if $\Delta(\tilde{\alpha},\tilde{\beta}) \leq \varepsilon$, and reject it otherwise. Therefore, given $N_\mathrm{sbi}$ such tries with discrepancies $\Delta_j(\tilde{\alpha},\tilde{\beta})$ ($1 \leq j \leq N_\mathrm{sbi}$), the resulting SBI posterior satisfies
\begin{equation}
\p_\mathrm{SBI}(\alpha,\beta|\tilde{\alpha}^\mathrm{obs}, \tilde{\beta}^\mathrm{obs}, \varepsilon) \propto \sum_{j=1}^{N_\mathrm{sbi}} \textbf{I}_{\left[ 0,\varepsilon \right]} \left[ \Delta_j(\tilde{\alpha},\tilde{\beta}) \right],
\label{eq:SBI_posterior}
\end{equation}
where $\textbf{I}_{\left[ 0,\varepsilon \right]}$ is the indicator function of the interval $\left[ 0,\varepsilon \right]$.

\section{Field-level inference via data assimilation}
\label{sec:DA}

\begin{figure*}
	\begin{center}
    	\includegraphics[width=\textwidth]{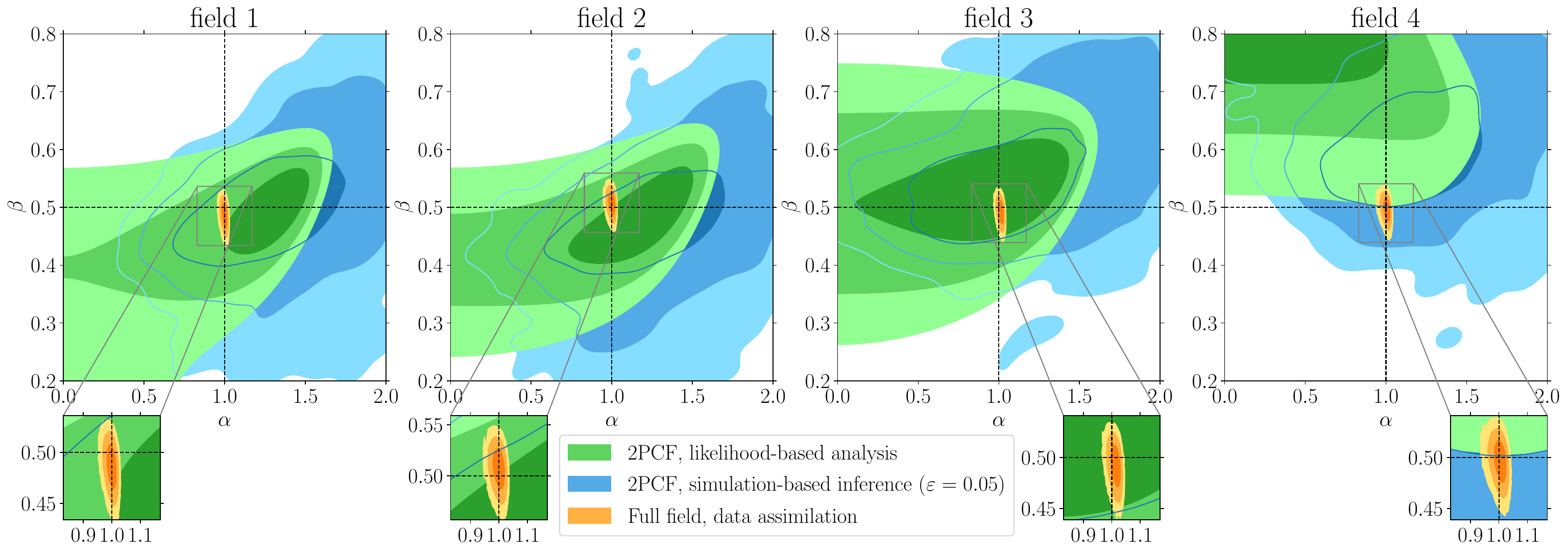}
    \end{center}
    \vspace*{-15pt}
    \caption{
    Comparison of different analysis techniques for four different log-normal fields. The posteriors are obtained using a likelihood-based analysis of the 2PCF (equation \eqref{eq:LBA_posterior}, green), simulation-based inference using the 2PCF (equation \eqref{eq:SBI_posterior}, blue), and data assimilation at the field level (equation \eqref{eq:DA_posterior}, orange). $1$-, $2$- and $3\sigma$ contours are shown. Ground truth values are $\alpha=1$ and $\beta=0.5$.
    }
    \label{fig:4fields}
\end{figure*}

The third and last technique we consider is a field-level analysis, where we infer $\alpha$ and $\beta$ from the full observed field $\bof^\mathrm{obs}$, rather than its estimated 2PCF. Field-level inference uses all of the data to hand. This is done by sampling all the variables appearing above $\bof$ in the BHM (i.e. $\alpha$, $\beta$ and $\bog$), conditional on $\bof^\mathrm{obs}$. This means that the $N_\mathrm{pix}$ variables of $\bog$, which are latent variables of the problem in sections \ref{sec:LBA} and \ref{sec:LFI}, now become target parameters of the problem. The full joint posterior satisfies
\begin{align}
\p(\alpha,\beta,\bog|\bof^\mathrm{obs}) & \propto \p(\bof^\mathrm{obs}|\bog,\alpha,\beta) \p(\bog,\alpha,\beta) \label{eq:full_joint_posterior} \\
& = \p(\bof^\mathrm{obs}|\bog,\alpha) \p(\bog|\beta) \p(\beta) \p(\alpha). \nonumber
\end{align}
The first term is given by equation \eqref{eq:p_f_LN}, the second by equation \eqref{eq:p_f_G}, and the third and fourth are the uniform priors. Only the probabilistic equations of the LN model (those given in section \ref{sec:Model}) appear here, without any additional compression or statistical assumption. Therefore, from a mathematical point of view, field-level inference provides the exact solution to the inference problem. 

Given its high dimensionality, sampling from the posterior written down in equation \eqref{eq:full_joint_posterior} requires advanced statistical techniques such as Hamiltonian Monte Carlo \citep{Duane1987} or more recent variants such as a No-U Turn Sampler \citep[NUTS,][]{HoffmanGelman2011}. We refer to such techniques as data assimilation techniques, as they permit assimilation of the full observed field values $\bof^\mathrm{obs}$ into the field-level LN model (equations \eqref{eq:p_f_G}--\eqref{eq:p_f_LN}). Once samples of $\p(\alpha,\beta,g|\bof^\mathrm{obs})$ are obtained, the DA posterior on $\alpha$ and $\beta$ is simply given by marginalising over the field values $\bog$,
\begin{equation}
\p_\mathrm{DA}(\alpha,\beta|\bof^\mathrm{obs}) = \int \p(\alpha,\beta,\bog|\bof^\mathrm{obs}) \, \drm \bog,
\label{eq:DA_posterior}
\end{equation}
which is trivially obtained from the sampled values of $\alpha$ and $\beta$.

\section{Numerical results}
\label{sec:Results}

\begin{figure}
	\begin{center}
    	\includegraphics[width=0.9\columnwidth]{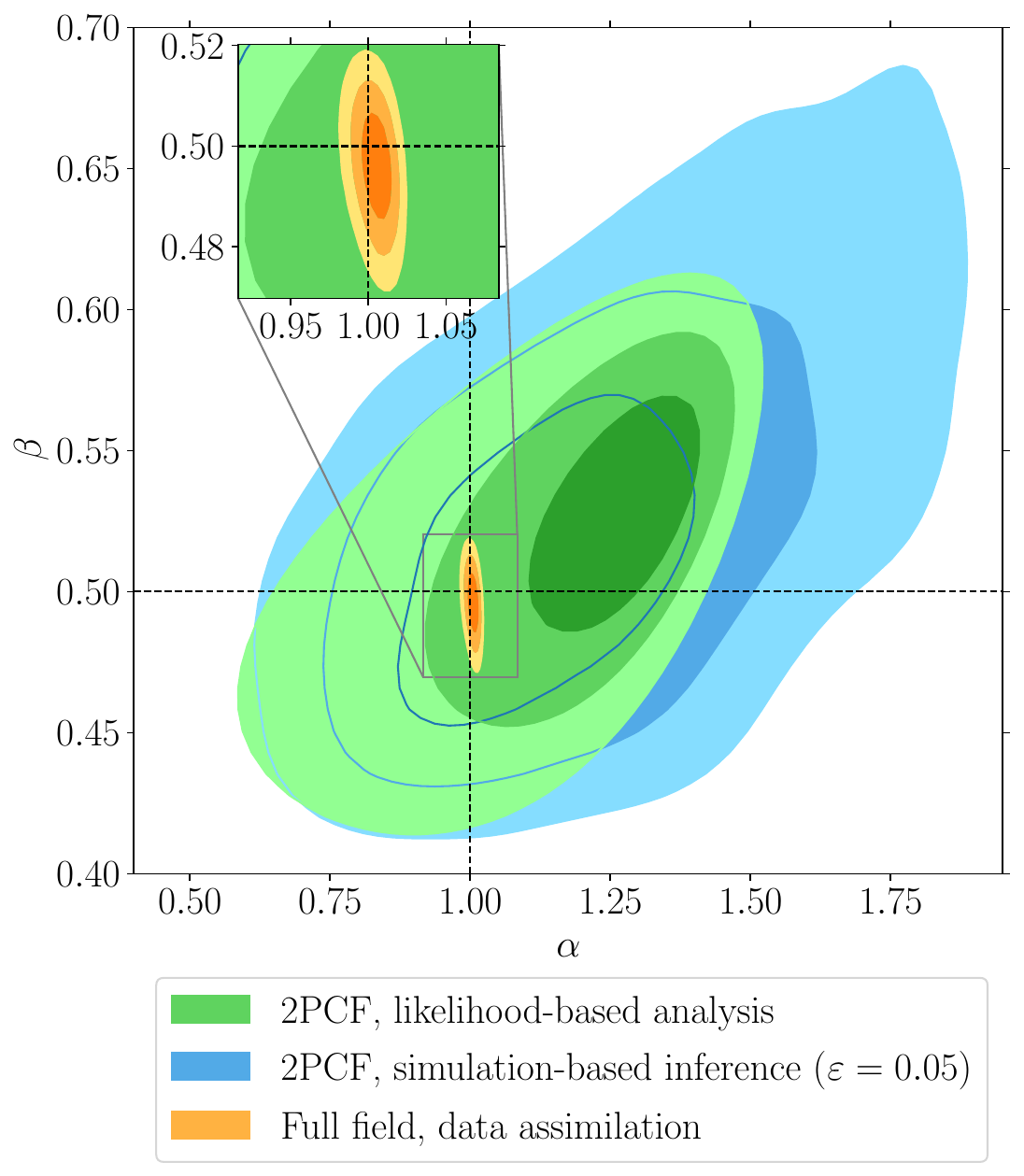}
    \end{center}
    \vspace*{-15pt}
    \caption{
    The posterior from the set of four independent log-normal fields. Using the same information (the 2PCF), simulation-based inference (blue) gives more accurate results than a likelihood-based analysis (green), which presents a bias of $\sim 2\sigma$. Using all the data to hand, field-level inference achieves unbiased and far more precise results.
    }
    \label{fig:4fields_comb}
\end{figure}

\begin{figure}
	\begin{center}
    	\includegraphics[width=0.88\columnwidth]{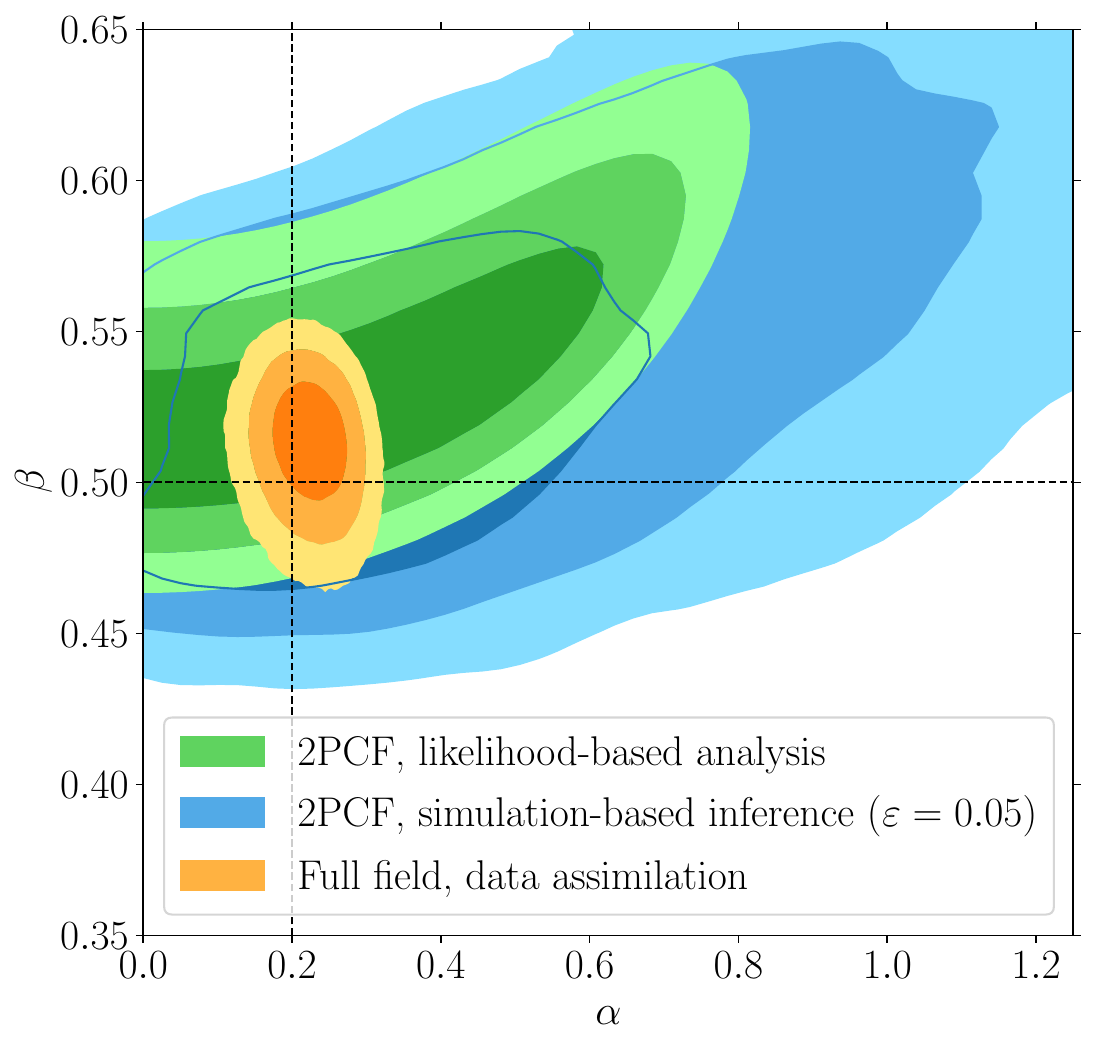}
    \end{center}
    \vspace*{-15pt}
    \caption{
    As in figure~\ref{fig:4fields}, except for a single near-Gaussian field with $\alpha=0.2$. Note that even for this almost Gaussian field, the 2PCF does not capture all the information; both a likelihood-based analysis and simulation-based inference fare well for the parameter $\beta$, but field-level inference is the only technique that gives precise and accurate results for $\alpha$.
    } 
    \label{fig:lowalpha}
\end{figure}

In this section we present some numerical results of our comparative study. 

We first analyse four different LN field realisations with ground truth values $\alpha = 1$ and $\beta = 0.5$ (see appendix \ref{apx:Fields analysed}). Corresponding posteriors on $\alpha$ and $\beta$, obtained with the three different methods, are shown in figure~\ref{fig:4fields}. For the LBA of $\boxi$, the posteriors $\p_\mathrm{LBA}(\alpha,\beta|\boxi^\mathrm{obs})$ (equation \eqref{eq:LBA_posterior}) are evaluated on a regular grid of $80 \times 80$ points and shown by the green contours. We used $N_\mathrm{sims} = 500$ simulations run at the ground truth values of $\alpha$ and $\beta$, well above the number of non-empty $r$-bins ($N_\xi=68$), so as to get an accurate estimate of $\boSig$ (see appendix \ref{apx:Simulations of the log-normal 2PCF}). For SBI, we use the ground truth as fiducial point ($\alpha_* = 1$, $\beta_* = 0.5$) to compress $\boxi$. We generated a large set of $N_\mathrm{sbi} \gtrsim 600,000$ simulations. Such a pool size is sufficient to ensure that at least $3,000$ samples are accepted when doing likelihood-free rejection sampling with $\varepsilon = 0.05$, for each of the four fields. The value of $\varepsilon = 0.05$ was chosen as a compromise between reducing the marginal variance of $\alpha$ and $\beta$ (by decreasing $\varepsilon$), and keeping a sensible acceptance rate (which goes to zero as $\varepsilon \rightarrow 0$, see appendix \ref{apx:Varying the threshold in SBI}). The SBI posteriors $\p_\mathrm{SBI}(\alpha,\beta|\tilde{\alpha}^\mathrm{obs}, \tilde{\beta}^\mathrm{obs}, \varepsilon)$ (equation \eqref{eq:SBI_posterior}) are shown in blue. For DA, we choose a NUTS initialised with automatic differentiation variational inference \citep[ADVI,][]{ADVI}. During the initialisation phase, ADVI automatically determines a variational family for the posterior and optimises the variational objective. Quantities required for NUTS, such as gradients of the data model with respect to the parameters ($\alpha$, $\beta$, $\bog$) and the mass matrix are also computed automatically. For each of the four fields $\bof^\mathrm{obs}$, we run $16$ independent chains. We determine the length of the residual burn-in phase (usually a few thousand samples), after ADVI initialisation, by examining trace plots of the parameters. We drop this burn-in phase, and obtain samples of the high-density regions of the posteriors. We enforce sufficient convergence by making sure that, for each of the $402$ parameters ($\alpha$, $\beta$ and the $20^2$ field values of $\bog$), the Gelman-Rubin statistic $\hat{R}$ is smaller than $1.05$. After marginalising over $\bog$, the final posteriors $\p_\mathrm{DA}(\alpha,\beta|\bof^\mathrm{obs})$ (equation \eqref{eq:DA_posterior}) are shown in orange in figure~\ref{fig:4fields}. We checked that the marginal standard deviations of $\alpha$ and $\beta$, obtained via LBA, SBI, and DA, are consistent with Cramér-Rao bounds, which can be computed analytically for a 2PCF analysis (see appendix \ref{apx:Fisher forecast for the 2PCF analysis}) and for a noise-free field-level analysis (see appendix \ref{apx:Fisher forecast for a full-field analysis}).

Since the four fields analysed in figure~\ref{fig:4fields} are independent, the final posterior for each method is obtained by multiplying the four likelihoods, and is shown in figure~\ref{fig:4fields_comb}. From figures~\ref{fig:4fields} and \ref{fig:4fields_comb}, it appears that the incorrect assumptions made in section \ref{sec:LBA} for the LBA of the 2PCF can cause strong biases, particularly when the field contains outlying values (e.g. field 4, see figure \ref{fig:fields}). Combining different fields does not entirely solve the issue. On the contrary, SBI using the same 2PCF provides accurate results, at a comparable level of precision. As the exact mathematical solution to the problem, field-level inference via DA achieves unbiased and far more precise results. The value of $\alpha = 1$ is rather large, so one might expect the 2PCF to lose information, but the relative performance of the field-level inference is strikingly good.

In the limit $\alpha\rightarrow 0$, the field becomes Gaussian, so one would expect the 2PCF (closer to a sufficient statistical summary of the field), to fare better. We investigated this hypothesis for a weakly non-Gaussian field with $\alpha=0.2$ (field 5 in figure \ref{fig:fields}), repeating the same treatment as described above. The results are shown in figure~\ref{fig:lowalpha}. In this regime, the 2PCF does capture well the spatial correlations of the field, as demonstrated by a comparable marginal variance on $\beta$ for the three methods. However, field-level inference remains the only technique that gives precise results for the non-Gaussianity parameter $\alpha$. This result can be interpreted from the model: as the LN transform (equation \eqref{eq:LN}) is a pixel-by-pixel operation, $\alpha$ is essentially one-point information, which is not captured well by the 2PCF, which is constant up to small corrections of order $\alpha^2$ when $\alpha \rightarrow 0$, as is readily seen from equation \eqref{eq:xi_LN}. Consistently, from equations \eqref{eq:dxi_LN_dalpha} and \eqref{eq:dxi_LN_dbeta}, it appears that $\partial \xi_\mathrm{LN}/\partial \alpha$ goes more rapidly to zero than $\partial \xi_\mathrm{LN}/\partial \beta$ as $r$ increases.

\section{Discussion and summary}
\label{sec:Discussion}

In this letter, we have compared the posteriors obtained from applying three data analysis techniques to log-normal fields. The log-normal field is occasionally used as an approximation to the late-time cosmological density field since its one-point distribution agrees approximately with that of the evolved matter density field, and by choice of the Gaussian two-point function, it can be made to agree at two-point level as well. The non-Gaussianity parameter $\alpha$ is a proxy for non-linear growth. The three techniques we have investigated are (\textit{i}) likelihood-based analysis: the standard technique of using correlation functions, with an assumption that these summary statistics are Gaussian-distributed; (\textit{ii}) simulation-based inference, or likelihood-free inference, where (optimally) compressed arbitrary summary statistics are used without statistical assumptions about their distribution; and (\textit{iii}) field-level inference, where the BHM is inferred in a mathematically exact way via data assimilation, and all the data are used. We find that for fields with significant non-Gaussianity, field-level inference gives far more precise and accurate posteriors than the standard LBA method, which can sometimes give highly inaccurate answers dependent on the data realisation. It is also far more precise than SBI using the 2PCF. Even for nearly Gaussian fields, field-level inference outperforms both in precision. 

We conclude that field-level inference should always be preferred to correlation functions when it is possible, particularly for non-Gaussian fields for which it gives the largest gain in precision. With specialised software, field-level inference with tens of millions of parameters is possible (e.g. \citealp{Lavaux2019} for galaxy clustering, \citealp{Porqueres2021} for weak lensing). When field-level inference is not possible for computational considerations, simulation-based inference gives more accurate answers than the standard likelihood-based analysis, which must arise from a breakdown of the Gaussian likelihood assumption (see appendix \ref{apx:Bias in LBA}, where we investigate the influence of our setup parameters on the accuracy of LBA).

The standard approach in weak lensing uses summary statistics assuming a Gaussian likelihood function, but this assumption is not strictly correct \citep{SellentinHeavens2018}. For correlation functions, \citet{Hartlap2009} and \citet{Sellentin2018} claim significant errors in posterior values. \citet{Lin2020} used principal component analysis to decorrelate the correlation function, showing small biases, but making a further assumption of independence. For pseudo-power spectra, a Gaussian likelihood may be accurate enough \citep{Taylor2019,Upham2021}, but the method needs the covariance matrix, which may be hard to calculate because of non-Gaussianities and super-sample covariance. Generally, likelihood-based analyses suffer from the difficulty of supplying an accurate covariance matrix, for which the parameter dependence, non-Gaussianity of the underlying field, and inclusion of super-sample covariance are challenging.

We note that there are prospects for improving the precision of simulation-based inference results by including more summary statistics, such as the field skewness, its three-point function, or summaries defined via machine learning. This is also the case for likelihood-based inference, but the problems of supplying an accurate covariance matrix, and the assumption of a Gaussian likelihood, still remain.

The code and data underlying this letter are publicly available on Github, at \url{https://github.com/florent-leclercq/correlations_vs_field}.

\section*{Acknowledgements}

We thank Andrew Jaffe, Guilhem Lavaux and Benjamin Wandelt for useful discussions. This work made use of the ELFI \citep{Lintusaari2017b}, pyDELFI \citep{Alsing2019} and pyMC3 \citep{pymc3} codes. This work was done within the \href{https://aquila-consortium.org}{Aquila Consortium}.

\section*{References}
\bibliography{biblio}

\onecolumngrid
\appendix

\section{Gradients of the log-normal 2PCF}
\label{apx:Gradients of the log-normal 2PCF}

Analytic expressions for the gradients of the 2PCF of LN fields (equation \eqref{eq:xi_LN}) are used for data compression within SBI analyses. They are given by
\begin{align}
\frac{\partial \xi_\mathrm{LN}}{\partial \alpha} & = \frac{2}{\alpha} \exp\left(-\frac{1}{4} \frac{r^2}{\beta^2}\right) \exp\left[\alpha^2\exp\left(-\frac{1}{4} \frac{r^2}{\beta^2}\right) \right] - \frac{2}{\alpha^3} \left\{ \exp\left[\alpha^2 \exp\left(-\frac{1}{4} \frac{r^2}{\beta^2}\right) \right] - 1 \right\}, \label{eq:dxi_LN_dalpha}\\
\frac{\partial \xi_\mathrm{LN}}{\partial \beta} & = \frac{1}{2}\frac{r^2}{\beta^3} \exp\left(-\frac{1}{4} \frac{r^2}{\beta^2}\right) \exp\left[\alpha^2 \exp\left(-\frac{1}{4} \frac{r^2}{\beta^2}\right) \right]. \label{eq:dxi_LN_dbeta}
\end{align}

\section{Fields analysed}
\label{apx:Fields analysed}
\setcounter{figure}{0}
\renewcommand\thefigure{\thesection.\arabic{figure}}    

\begin{figure*}
	\begin{center}
    	\includegraphics[width=\textwidth]{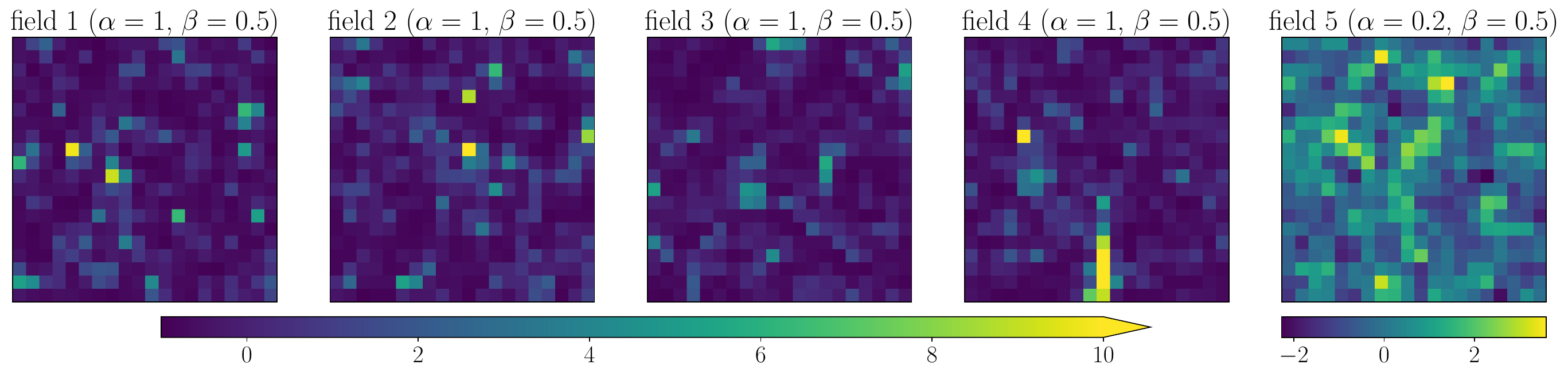}
    \end{center}
    \vspace*{-15pt}
    \caption{
	The different LN fields analysed in this paper, with ground truth values of $\alpha$ and $\beta$ indicated above the panels.
    }
    \vspace*{-10pt}
    \label{fig:fields}
\end{figure*}

Figure \ref{fig:fields} shows the different LN fields analysed in section \ref{sec:Results}. The first four fields have ground truth values $\alpha=1$ and $\beta=0.5$. Field 5 is a weakly non-Gaussian field with ground truth values $\alpha=0.2$ and $\beta=0.5$.

\section{Simulations of the log-normal 2PCF}
\label{apx:Simulations of the log-normal 2PCF}
\setcounter{figure}{0}
\renewcommand\thefigure{\thesection.\arabic{figure}}    

\begin{figure*}
	\begin{center}
    	\includegraphics[width=\textwidth]{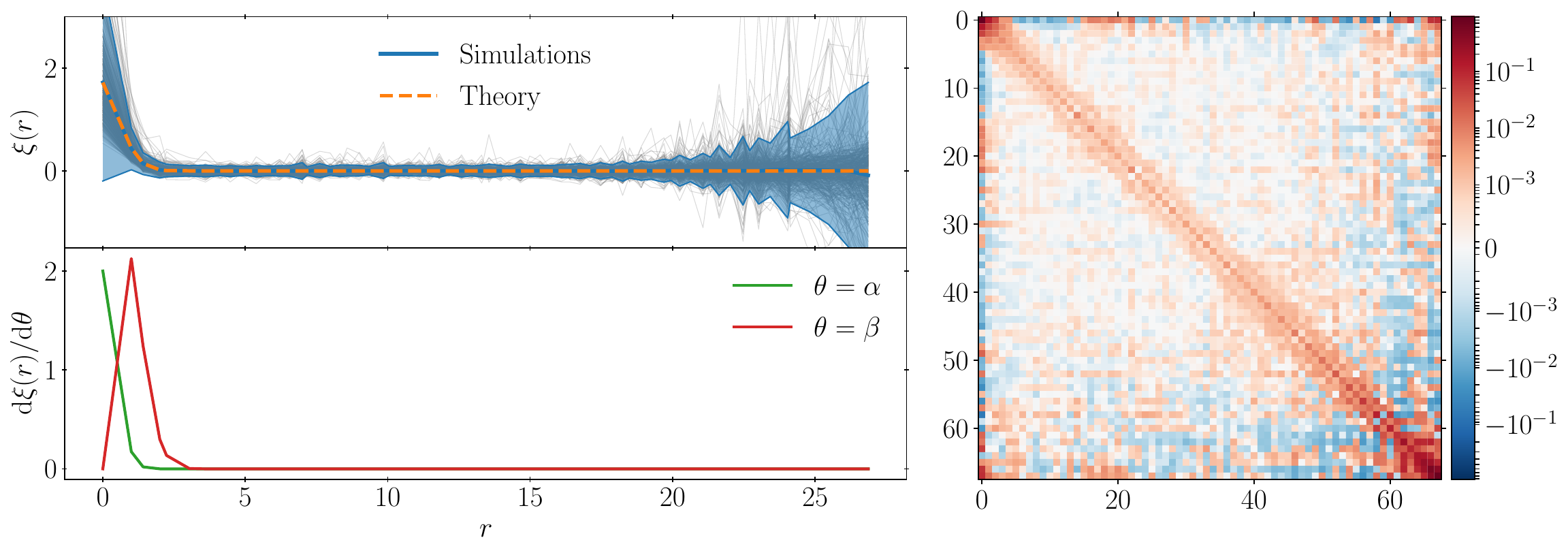}
    \end{center}
    \vspace*{-15pt}
    \caption{
    Simulations of the 2PCF of LN fields run at ground truth values $\alpha = 1$, $\beta = 0.5$. The top left-hand panel shows the simulated 2PCF as grey lines, their sample mean $\bar{\boxi}$ and corresponding $2\sigma$ uncertainty as the blue line and blue shaded region, and the theoretical expectation (equation \eqref{eq:xi_LN}) as the dashed orange line. The bottom left-hand panel shows the analytic derivatives of the LN 2PCF with respect to $\alpha$ and $\beta$ (equations \eqref{eq:dxi_LN_dalpha} and \eqref{eq:dxi_LN_dbeta}). The right-hand panel shows the estimated covariance matrix $\boSighat$ (equation \eqref{eq:covariance_estimator}), used both for LBA (equation \eqref{eq:LBA_posterior}) and data compression for SBI (equations \eqref{eq:compression_mle} and \eqref{eq:compression_fisher}).
    }
    \label{fig:2PCF_sims}
    \vspace*{-5pt}
\end{figure*}

For use in LBA and SBI, we ran $N_\mathrm{sims} = 500$ LN simulations at ground truth values $\alpha = 1$, $\beta = 0.5$. Their 2PCF $\boxi$, using $N_\xi = 68$ $r$-bins, are plotted as grey lines in the top left-hand panel of figure \ref{fig:2PCF_sims}. The covariance matrix $\boSighat$, estimated from these simulations using the estimator given in equation \eqref{eq:covariance_estimator}, is shown in the right-hand panel of figure \ref{fig:2PCF_sims}. The sample mean $\bar{\boxi}$ and corresponding $2\sigma$ uncertainty, i.e. $2\sqrt{\mathrm{diag}(\boSighat)}$, are plotted in blue in the top left-hand panel of figure \ref{fig:2PCF_sims}, along with the theoretical expectation given by equation \eqref{eq:xi_LN} (dashed orange line). Analytic gradients of $\xi_\mathrm{LN}$, given by equations \eqref{eq:dxi_LN_dalpha} and \eqref{eq:dxi_LN_dbeta}, are plotted in the bottom left-hand panel of figure \ref{fig:2PCF_sims}.

\section{Varying the threshold in SBI}
\label{apx:Varying the threshold in SBI}
\setcounter{figure}{0}
\renewcommand\thefigure{\thesection.\arabic{figure}}    

\begin{figure*}
	\begin{center}
    	\includegraphics[width=\textwidth]{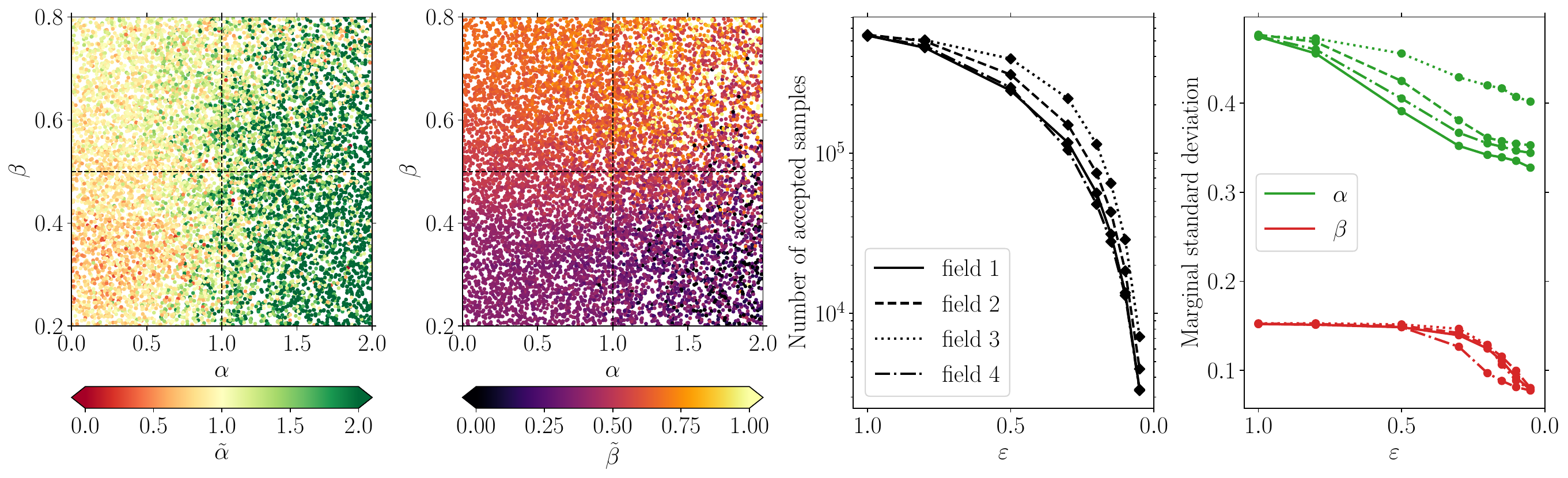}
    \end{center}
    \vspace*{-15pt}
    \caption{
    SBI analysis of the 2PCF of LN fields. Fiducial values for data compression are $\alpha_* = 1$ and $\beta_* = 0.5$. From left to right, the first two panels show the compressed data $\tilde{\alpha}$ and $\tilde{\beta}$ in some of our simulations. The third panel shows the number of accepted samples as a function of the threshold $\varepsilon \in \left\lbrace 1.0,0.80,0.50,0.30,0.20,0.15,0.10,0.05 \right\rbrace$ for four different fields, as indicated by different line styles. The last panel shows the marginal standard deviation of the inferred parameters $\alpha$ and $\beta$ as a function of $\varepsilon$.
    }
    \vspace*{-5pt}
    \label{fig:SBI_sims}
\end{figure*}

\begin{figure*}
	\begin{center}
    	\includegraphics[width=\textwidth]{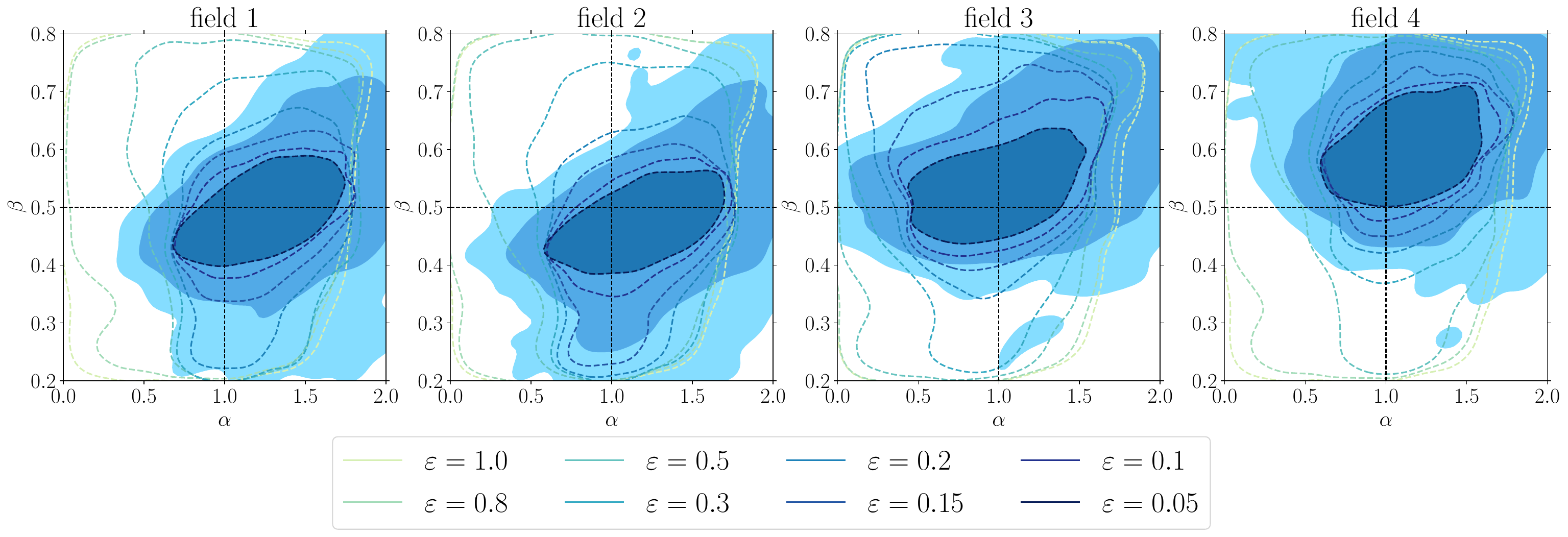}
    \end{center}
    \vspace*{-15pt}
    \caption{
    Effect of varying $\varepsilon$ on the SBI posterior. The dashed lines show the $1\sigma$ contours of $\p_\mathrm{SBI}(\alpha,\beta|\tilde{\alpha}^\mathrm{obs}, \tilde{\beta}^\mathrm{obs}, \varepsilon)$ for different values of $\varepsilon$ as indicated in the caption. The filled contours show the $1$-, $2$- and $3\sigma$ contours of $\p_\mathrm{SBI}(\alpha,\beta|\tilde{\alpha}^\mathrm{obs}, \tilde{\beta}^\mathrm{obs}, \varepsilon)$ for $\varepsilon = 0.05$, as in figure \ref{fig:4fields}.
    }
    \label{fig:epsilons}
\end{figure*}

In this appendix, we discuss the effect of varying the threshold $\varepsilon$ on the SBI posterior $\p_\mathrm{SBI}(\alpha,\beta|\tilde{\alpha}^\mathrm{obs}, \tilde{\beta}^\mathrm{obs}, \varepsilon)$ (equation \eqref{eq:SBI_posterior}). In figure \ref{fig:SBI_sims}, the two leftmost panels represent the values of $\tilde{\alpha}$ and $\tilde{\beta}$ in a subset of the $N_\mathrm{sbi} \gtrsim 600,000$ simulations used in section \ref{sec:Results}. Fiducial values used for data compression are $(\alpha_*,\beta_*) = (1, 0.5)$ and the observed compressed data are $(\tilde{\alpha}^\mathrm{obs}, \tilde{\beta}^\mathrm{obs}) = (1.589, 0.503)$, $(1.449, 0.475)$, $(1.248, 0.551)$, $(1.528, 0.648)$ for fields 1, 2, 3, and 4, respectively (note that a different set of simulations, compressed with fiducial values $(\alpha_*,\beta_*) = (0.2, 0.5)$, is used to analyse field 5, which has $(\tilde{\alpha}^\mathrm{obs}, \tilde{\beta}^\mathrm{obs}) = (0.441, 0.531)$).

The two rightmost panels of figure \ref{fig:SBI_sims} show diagnostics of likelihood-free rejection sampling with different values of the threshold $\varepsilon$, for each of the four fields. The third panel shows the number of accepted samples as a function of $\varepsilon$, and the fourth panel shows the marginal standard deviation of $\alpha$ and $\beta$, obtained from the joint posterior $\p_\mathrm{SBI}(\alpha,\beta|\tilde{\alpha}^\mathrm{obs}, \tilde{\beta}^\mathrm{obs}, \varepsilon)$, as a function of $\varepsilon$. As expected, the number of accepted samples and the marginal standard deviation of parameters both decrease as a function of $\varepsilon$. As discussed in section \ref{sec:Results}, we chose $\varepsilon = 0.05$ as a compromise between the acceptance ratio and the reduction of the marginal standard deviation of parameters. This threshold guarantees that at least $3,000$ samples are accepted to estimate the SBI posterior, for each of the four fields. 

Figure \ref{fig:epsilons} shows the effect of varying $\varepsilon$ on the two-dimensional SBI posterior. There, the $1\sigma$ contour of $\p_\mathrm{SBI}(\alpha,\beta|\tilde{\alpha}^\mathrm{obs}, \tilde{\beta}^\mathrm{obs}, \varepsilon)$ for $\varepsilon \in \left\lbrace 1.0,0.80,0.50,0.30,0.20,0.15,0.10,0.05 \right\rbrace$ is represented, along with the final contours, corresponding to $\varepsilon = 0.05$.

\section{Fisher forecast for the 2PCF analysis}
\label{apx:Fisher forecast for the 2PCF analysis}

In this appendix, we present a Fisher forecast for the LBA of the 2PCF of LN fields. The Fisher information matrix $\Fxi$ is defined by its elements
\begin{equation}
F^\xi_{\alpha \alpha} \equiv - \left\langle \frac{\partial^2 \ln \p(\boxi|\alpha,\beta)}{\partial \alpha^2} \right\rangle ; \quad F^\xi_{\alpha \beta} \equiv - \left\langle \frac{\partial^2 \ln \p(\boxi|\alpha,\beta)}{\partial \alpha \partial \beta} \right\rangle ; \quad F^\xi_\mathrm{\beta\beta} \equiv - \left\langle \frac{\partial^2 \ln \p(\boxi|\alpha,\beta)}{\partial \beta^2} \right\rangle .
\label{eq:def_Fisher_2PCF}
\end{equation}
Assuming that $\boxi$ follows a multivariate Gaussian distribution with mean $\boxi_\mathrm{LN}(\alpha,\beta)$ and parameter-independent covariance matrix $\boSig$ (equation \eqref{eq:LBA_Gaussian_assumption}), the elements of $\Fxi$ are given by \citep{Tegmark1997}
\begin{equation}
F^{\xi}_{ij} = \frac{1}{2} \tr(\boSig^{-1} \Mij),
\end{equation}
where $\Mij$ is the $N_\xi \times N_\xi$ matrix defined by $\Mij \equiv \boxi_{\mathrm{LN},i}\boxi_{\mathrm{LN},j}^\intercal + \boxi_{\mathrm{LN},j}\boxi_{\mathrm{LN},i}^\intercal$, for $i$, $j$ $\in \{\alpha , \beta\}$. The comma notation denotes derivatives, for instance $\boxi_{\mathrm{LN},\alpha} \equiv \partial \boxi_\mathrm{LN} / \partial \alpha$. An equivalent expression is given by equation \eqref{eq:fisher_definition_gaussian}, i.e.
\begin{equation}
\Fxi = \nabla \boxi_\mathrm{LN}^\intercal \boSig^{-1} \nabla \boxi_\mathrm{LN},
\label{eq:Fisher_2PCF_Gaussian}
\end{equation}
where $\nabla \boxi_\mathrm{LN}^\intercal$ is the $2 \times N_\xi$ matrix defined by
\begin{equation}
\nabla \boxi_\mathrm{LN}^\intercal \equiv \begin{pmatrix}
\boxi_{\mathrm{LN},\alpha}^\intercal \\
\boxi_{\mathrm{LN},\beta}^\intercal
\end{pmatrix}.
\end{equation}
A straightforward (but slightly incorrect) way to estimate $\Fxi$ would be to replace the unknown covariance matrix $\boSig$ by the estimated $\boSighat$ in equation \eqref{eq:Fisher_2PCF_Gaussian}. Taking into account the modification of the Gaussian distribution to a $t$-distribution, a more accurate calculation of the Fisher information matrix has been performed by \citet{SellentinHeavens2017}. For a parameter-independent covariance matrix, the correction reduces to a prefactor: an estimator of the inverse Fisher information matrix $(\Fxi)^{-1}$ is given by
\begin{equation}
\widehat{(\Fxi)^{-1}} \equiv \frac{N_\mathrm{sims} - 1}{N_\mathrm{sims} - p + N_\xi -1} (\nabla \boxi_\mathrm{LN}^\intercal \boSighat^{-1} \nabla \boxi_\mathrm{LN})^{-1},
\label{eq:Fisher_2PCF_tdistrib}
\end{equation}
where $p=2$ is the number of model parameters. The Cramér-Rao bounds on $\alpha$ and $\beta$ are $\Delta^\xi \alpha \equiv \sqrt{(F^{\xi})^{-1}_{\alpha\alpha}}$ and $\Delta^\xi \beta \equiv \sqrt{(F^{\xi})^{-1}_{\beta\beta}}$, where the $(F^{\xi})^{-1}_{ij}$ are the elements of the matrix defined by equation \eqref{eq:Fisher_2PCF_tdistrib}.

Using our estimated covariance matrices $\boSighat$, we find the following numerical values for the Cramér-Rao bounds:
\begin{align}
& \Delta^\xi \alpha \approx 0.343,~\Delta^\xi \beta \approx 0.058 \quad \mathrm{for} \quad \alpha =1,~\beta = 0.5, \quad \mathrm{and} \\
& \Delta^\xi \alpha \approx 0.465 ,~\Delta^\xi \beta \approx 0.029 \quad \mathrm{for} \quad \alpha =0.2,~\beta = 0.5.
\end{align}
For the fields analysed in this work, the realised marginal errors on $\alpha$ and $\beta$ are, for LBA,
\begin{align}
& \sigma_\alpha \approx 0.355,~\sigma_\beta \approx 0.064 \quad \mathrm{(field~1)}, \\
& \sigma_\alpha \approx 0.388,~\sigma_\beta \approx 0.060 \quad \mathrm{(field~2)}, \\
& \sigma_\alpha \approx 0.394,~\sigma_\beta \approx 0.067 \quad \mathrm{(field~3)}, \\
& \sigma_\alpha \approx 0.352,~\sigma_\beta \approx 0.052 \quad \mathrm{(field~4)}, \\
& \sigma_\alpha \approx 0.180,~\sigma_\beta \approx 0.024 \quad \mathrm{(field~5)};
\end{align}
and for SBI,
\begin{align}
& \sigma_\alpha \approx 0.328,~\sigma_\beta \approx 0.078 \quad \mathrm{(field~1)}, \\
& \sigma_\alpha \approx 0.353,~\sigma_\beta \approx 0.080 \quad \mathrm{(field~2)}, \\
& \sigma_\alpha \approx 0.402,~\sigma_\beta \approx 0.078 \quad \mathrm{(field~3)}, \\
& \sigma_\alpha \approx 0.344,~\sigma_\beta \approx 0.077 \quad \mathrm{(field~4)}, \\
& \sigma_\alpha \approx 0.330,~\sigma_\beta \approx 0.050 \quad \mathrm{(field~5)}.
\end{align}
Although it is difficult to be precise since LBA and SBI posteriors are quite variable, we note that the above numbers are in rough agreement with the Cramér-Rao bounds for the 2PCF analysis. In particular, $\sigma_\alpha \gtrsim \Delta^\xi \alpha$ and $\sigma_\beta \gtrsim \Delta^\xi \beta$ hold for most realisations.\footnote{The information inequality seems to be violated for field 5, giving $\sigma_\alpha < \Delta^\xi \alpha$. Nevertheless, the small marginal standard deviation of $\alpha$ includes the effect of the sharp prior boundary at $\alpha=0$, visible in figure \ref{fig:lowalpha}, which is not accounted for in the Fisher calculation, based only on the likelihood (see equation \eqref{eq:def_Fisher_2PCF}).}

\section{Fisher forecast for a full-field analysis}
\label{apx:Fisher forecast for a full-field analysis}

In this appendix, we calculate the Fisher information matrix for a field-level analysis of LN fields, and compare the Cramér-Rao bounds to the results obtained via DA in section \ref{sec:Results}.

\subsection{Calculation of the Fisher matrix}

To make the calculation tractable, we simplify the BHM described in section \ref{sec:Model}: we work in a noise-free setting, in which $\bof$ and $\bog$ are deterministically linked by (see equations \eqref{eq:LN} and \eqref{eq:p_f_LN})
\begin{equation}
\bof = \frac{1}{\alpha}\left[ \exp\left(\alpha \bog - \frac{\alpha^2}{2}\right) -1 \right], \quad \bog=\frac{1}{\alpha} \left[ \ln(\alpha \bof +1) + \frac{\alpha^2}{2} \right].
\label{eq:LN_mapping_noisefree}
\end{equation}
We have $\p(\bof|\alpha,\beta) \, \drm \bof = \p(\bog|\alpha, \beta) \, \drm \bog = \p(\bog|\beta) \, \drm \bog$. The Jacobian matrix has elements
\begin{equation}
\deriv{f_i}{g_j} = (\alpha f_i + 1) \delta_\mathrm{K}^{ij},
\end{equation}
for $i \in \llbracket 1, N_\mathrm{pix} \rrbracket$ and $j \in \llbracket 1, N_\mathrm{pix} \rrbracket$, where $\updelta_\mathrm{K}^{ij}$ denotes a Kronecker delta. Therefore, the sampling distribution for $\bof$ is given by
\begin{align}
\p(\bof|\alpha, \beta) & = \p\left\lbrace\left. \frac{1}{\alpha} \left[ \ln(\alpha \bof +1) + \frac{\alpha^2}{2} \right] \right| \beta \right\rbrace \times \frac{1}{\prod_i (\alpha f_i +1)} \nonumber\\
& = \frac{\exp\left\lbrace -\frac{1}{2\alpha^2} \left[ \ln(\alpha \bof +1) + \frac{\alpha^2}{2} \right]^\intercal \XiG^{-1} \left[ \ln(\alpha \bof +1) + \frac{\alpha^2}{2} \right] \right\rbrace}{\sqrt{|2\pi \XiG|}} \times \frac{1}{\prod_i (\alpha f_i +1)}.
\end{align}
The log-probability is
\begin{align}
\ln \p(\bof|\alpha, \beta) & = -\frac{1}{2} \ln |\XiG| -\frac{1}{2\alpha^2} \left[ \ln(\alpha \bof +1) + \frac{\alpha^2}{2} \right]^\intercal \XiG^{-1} \left[ \ln(\alpha \bof +1) + \frac{\alpha^2}{2} \right] - \sum_i \ln(\alpha f_i +1) + \mathrm{const.} \nonumber\\
& = -\frac{1}{2} \tr( \ln \XiG + \XiG^{-1} \boK ) - \sum_i \ln(\alpha f_i +1) + \mathrm{const.} \label{eq:log_likelihood_Fisher}
\end{align}
where we have used the well-known identity $\ln |\XiG| = \tr(\ln \XiG)$, and introduced the matrix
\begin{equation}
\boK \equiv \frac{1}{\alpha^2} \left[ \ln(\alpha \bof +1) + \frac{\alpha^2}{2} \right] \left[ \ln(\alpha \bof +1) + \frac{\alpha^2}{2} \right]^\intercal = \bog \bog^\intercal.
\end{equation}
We note that for fixed $\bof$ (or $\bog$), $\XiG$ depends only on $\beta$ and $\boK$ depends only on $\alpha$. By construction, $\left\langle \bog \right\rangle = 0$ and $\left\langle \boK \right\rangle = \int \boK \, \p(\bof|\alpha,\beta) \, \drm \bof = \int \bog \bog^\intercal \p(\bog|\beta) \, \drm \bog = \left\langle \bog \bog^\intercal \right\rangle = \XiG$. 

We start with some preliminary calculations. For any vector $\boj$ and any function $F$, we can write the Gaussian integral:
\begin{equation}
\left\langle \erm^{\boj^\intercal \bog} F(\bog) \right\rangle = \int F(\bog) \frac{\erm^{-\frac{1}{2} \bog^\intercal \XiG^{-1} \bog + \boj^\intercal \bog}}{\sqrt{|2\pi \XiG |}} \, \drm \bog = \erm^{\frac{1}{2} \boj^\intercal \XiG \boj} \int F(\boy + \XiG \boj) \frac{\erm^{-\frac{1}{2} \boy^\intercal \XiG^{-1} \boy}}{\sqrt{|2\pi \XiG|}} \, \drm \boy.
\label{eq:Gaussian_integral}
\end{equation}
From equation \eqref{eq:Gaussian_integral}, using respectively $j_k = \lambda \updelta_\mathrm{K}^{ik}$ and $F(\bog)=1$; $j_k = \lambda (\updelta_\mathrm{K}^{ik}+\updelta_\mathrm{K}^{jk})$ and $F(\bog)=1$; and $j_k = \lambda \updelta_\mathrm{K}^{ik}$ and $F(\bog) = g_j$, we deduce the following standard results: for any scalars $\lambda$ and $\mu$,
\begin{align}
\left\langle \erm^{\lambda \bog + \mu} \right\rangle & = \erm^{\frac{\lambda^2}{2} + \mu}; \label{eq:LN_lemma_1}\\
\left\langle \erm^{\lambda \bog} \left(\erm^{\lambda \bog}\right)^\intercal \right\rangle & = \erm^{\lambda^2} \exp(\lambda^2 \XiG); \label{eq:LN_lemma_2}\\
\left\langle \erm^{\lambda \bog} \bog^\intercal \right\rangle & = \erm^{\frac{\lambda^2}{2}} \lambda \XiG. \label{eq:LN_lemma_3}
\end{align}
From these relations, we recover $\left\langle \bof \right\rangle = 0$ and the covariance matrix of LN fields (equation \eqref{eq:xi_LN}):
\begin{equation}
\left\langle \bof \bof^\intercal \right\rangle = \frac{1}{\alpha^2} \left\langle \left( \erm^{\alpha \bog - \frac{\alpha^2}{2}} -1 \right) \left( \erm^{\alpha \bog- \frac{\alpha^2}{2}} -1 \right)^\intercal \right\rangle = \frac{1}{\alpha^2} \left[ \exp(\alpha^2 \XiG) - \boldsymbol{1}.\boldsymbol{1}^\intercal \right].
\label{eq:exp_f_f_T}
\end{equation}

Let us define $\boh \equiv \bof/(\alpha \bof + 1)$, $\boQ \equiv \left\langle \boh \boh^\intercal \right\rangle$, $\boR \equiv \left\langle \boh \bog^\intercal \right\rangle$, and $\boS \equiv \left\langle \boh^2 \bog^\intercal \right\rangle$. Using equation \eqref{eq:LN_lemma_1}, we get
\begin{equation}
\left\langle \boh \right\rangle = \frac{1}{\alpha} \left\langle 1 - \erm^{- \alpha \bog + \frac{\alpha^2}{2}} \right\rangle = \frac{1}{\alpha}\left(1 - \erm^{\alpha^2}\right),
\label{eq:expt_h}
\end{equation}
and using equations \eqref{eq:LN_lemma_1} and \eqref{eq:LN_lemma_2}, we get
\begin{align}
\boQ = \left\langle \boh \boh^\intercal \right\rangle & = \frac{1}{\alpha^2} \left\langle \left( 1 - \erm^{- \alpha \bog + \frac{\alpha^2}{2}} \right) \left( 1 - \erm^{- \alpha \bog + \frac{\alpha^2}{2}} \right)^\intercal \right\rangle \nonumber \\
& = \frac{1}{\alpha^2} \left[ \erm^{\alpha^2} \left\langle \erm^{-\alpha \bog} \left( \erm^{-\alpha \bog} \right)^\intercal \right\rangle - \left\langle \erm^{-\alpha \bog + \frac{\alpha^2}{2}}.\boldsymbol{1}^\intercal \right\rangle - \left\langle \boldsymbol{1}.\left(\erm^{-\alpha \bog + \frac{\alpha^2}{2}}\right)^\intercal \right\rangle + \boldsymbol{1}.\boldsymbol{1}^\intercal \right] \nonumber\\
& = \frac{1}{\alpha^2} \left[ \erm^{2\alpha^2} \exp(\alpha^2 \XiG) + (1-2\erm^{\alpha^2}) \boldsymbol{1}.\boldsymbol{1}^\intercal \right], \label{eq:Q_matrix}
\end{align}
from which (at $r=0$),
\begin{equation}
\left\langle \boh^2 \right\rangle = \frac{1}{\alpha^2} \left(\erm^{3\alpha^2} - 2\erm^{\alpha^2} + 1 \right).
\label{eq:expt_h2}
\end{equation}
Furthermore, using equation \eqref{eq:LN_lemma_3},
\begin{align}
\boR & = \left\langle \boh \bog^\intercal \right\rangle = \frac{1}{\alpha} \left\langle \left( 1 - \erm^{- \alpha \bog + \frac{\alpha^2}{2}} \right) \bog^\intercal \right\rangle = - \frac{\erm^{\frac{\alpha^2}{2}}}{\alpha} \left\langle \erm^{-\alpha \bog} \bog^\intercal \right\rangle = \erm^{\alpha^2} \XiG ; \label{eq:R_matrix}\\
\boS & = \left\langle \boh^2 \bog^\intercal \right\rangle = \frac{1}{\alpha^2} \left[ \left\langle \left( 1 -2\erm^{-\alpha \bog + \frac{\alpha^2}{2}} + \erm^{-2\alpha \bog + \alpha^2} \right) \bog^\intercal \right\rangle \right] = \frac{1}{\alpha^2} \left[ -2\erm^{\frac{\alpha^2}{2}} \left\langle \erm^{-\alpha \bog} \bog^\intercal \right\rangle + \erm^{\alpha^2} \left\langle \erm^{-2\alpha \bog} \bog^\intercal \right\rangle \right] \nonumber\\
& = \frac{2}{\alpha} \left(\erm^{\alpha^2} - \erm^{3\alpha^2} \right) \XiG. \label{eq:S_matrix}
\end{align}

We now compute $\left\langle \boK_{,\alpha} \right\rangle$ and $\left\langle \boK_{,\alpha\alpha} \right\rangle$. From the LN mapping (equation \eqref{eq:LN_mapping_noisefree}), we get 
\begin{equation}
\bog_{,\alpha} = -\frac{1}{\alpha^2} \ln(\alpha \bof +1) + \frac{\bof}{\alpha(\alpha \bof +1)} + \frac{1}{2} = -\frac{1}{\alpha} \bog + \frac{1}{\alpha} \boh + 1.
\end{equation}
Therefore, using equation \eqref{eq:R_matrix},
\begin{equation}
\left\langle \boK_{,\alpha} \right\rangle = \left\langle \bog_{,\alpha} \bog^\intercal + \bog \bog_{,\alpha}^\intercal \right\rangle = - \frac{2}{\alpha} \left\langle \bog \bog^\intercal \right\rangle + \frac{1}{\alpha} \left\langle \boh \bog^\intercal + \bog \boh^\intercal \right\rangle = - \frac{2}{\alpha} \XiG + \frac{2}{\alpha} \boR = \frac{2}{\alpha}(\erm^{\alpha^2}-1) \XiG. \label{eq:Ka}
\end{equation}
Moreover,
\begin{align}
\bog_{,\alpha} \bog_{,\alpha}^\intercal & = \frac{1}{\alpha^2} \bog \bog^\intercal + \frac{1}{\alpha^2} \boh \boh^\intercal + \boldsymbol{1}.\boldsymbol{1}^\intercal - \frac{1}{\alpha^2} (\boh \bog^\intercal + \bog \boh^\intercal) - \frac{1}{\alpha} (\bog.\boldsymbol{1}^\intercal + \boldsymbol{1}.\bog^\intercal) + \frac{1}{\alpha} (\boh.\boldsymbol{1}^\intercal + \boldsymbol{1}.\boh^\intercal); \\
\left\langle \bog_{,\alpha} \bog_{,\alpha}^\intercal \right\rangle & = \frac{1}{\alpha^2} \XiG + \frac{1}{\alpha^2} \boQ + \left[1+ \frac{2}{\alpha^2} \left( 1-\erm^{\alpha^2} \right) \right] \boldsymbol{1}.\boldsymbol{1}^\intercal - \frac{2}{\alpha^2} \boR, \label{eq:Kaa_first_term}
\end{align}
where we have used equation \eqref{eq:expt_h}. The second derivative of $\bog$ is
\begin{equation}
\bog_{,\alpha\alpha} = \frac{2}{\alpha^3} \ln(\alpha \bof +1) - \frac{1}{\alpha^2} \frac{\bof}{\alpha \bof +1}  - \frac{1}{\alpha^2} \frac{\bof}{\alpha \bof +1} - \frac{1}{\alpha} \frac{\bof^2}{(\alpha \bof +1)^2} = \frac{2}{\alpha^2} \bog - \frac{1}{\alpha} - \frac{2}{\alpha^2}\boh - \frac{1}{\alpha}\boh^2,
\end{equation}
which gives
\begin{equation}
\left\langle \bog_{,\alpha\alpha} \bog^\intercal + \bog \bog_{,\alpha\alpha}^\intercal \right\rangle = \frac{4}{\alpha^2} \left\langle \bog \bog^\intercal \right\rangle - \frac{2}{\alpha^2}  \left\langle \boh \bog^\intercal + \bog \boh^\intercal \right\rangle - \frac{1}{\alpha} \left\langle \boh^2 \bog^\intercal + \bog \boh^{2\intercal} \right\rangle = \frac{4}{\alpha^2} \XiG - \frac{4}{\alpha^2} \boR - \frac{2}{\alpha}\boS.
\label{eq:Kaa_second_term}
\end{equation}
Equations \eqref{eq:Kaa_first_term} and \eqref{eq:Kaa_second_term} and the expressions for $\boQ$, $\boR$ and $\boS$ (equations \eqref{eq:Q_matrix}, \eqref{eq:R_matrix} and \eqref{eq:S_matrix}) yield
\begin{align}
\left\langle \boK_{,\alpha\alpha} \right\rangle & = \left\langle \bog_{,\alpha\alpha} \bog^\intercal + \bog \bog_{,\alpha\alpha}^\intercal + 2 \bog_{,\alpha} \bog_{,\alpha}^\intercal \right\rangle = \frac{6}{\alpha^2} \XiG - \frac{8}{\alpha^2} \boR - \frac{2}{\alpha}\boS + \frac{2}{\alpha^2} \boQ + 2 \left[1+ \frac{2}{\alpha^2} \left( 1-\erm^{\alpha^2} \right) \right] \boldsymbol{1}.\boldsymbol{1}^\intercal \nonumber\\
& = \frac{2}{\alpha^2} \left(3 - 6\erm^{\alpha^2} + 2\erm^{3\alpha^2}\right) \XiG + \frac{2}{\alpha^4} \erm^{2\alpha^2} \exp(\alpha^2 \XiG) + 2\left[\frac{1}{\alpha^4} (1-2\erm^{\alpha^2}) + 1 + \frac{2}{\alpha^2} (1-\erm^{\alpha^2}) \right] \boldsymbol{1}.\boldsymbol{1}^\intercal. \label{eq:Kaa}
\end{align}

Finally, we compute the elements of the Fisher information matrix $\Ff$, defined by
\begin{equation}
F^f_{\alpha \alpha} \equiv - \left\langle \frac{\partial^2 \ln \p(\bof|\alpha,\beta)}{\partial \alpha^2} \right\rangle ; \quad F^f_{\alpha \beta} \equiv - \left\langle \frac{\partial^2 \ln \p(\bof|\alpha,\beta)}{\partial \alpha \partial \beta} \right\rangle ; \quad F^f_\mathrm{\beta\beta} \equiv - \left\langle \frac{\partial^2 \ln \p(\bof|\alpha,\beta)}{\partial \beta^2} \right\rangle .
\end{equation}
Starting from equation \eqref{eq:log_likelihood_Fisher},
\begin{equation}
\frac{\partial \ln \p(\bof|\alpha,\beta)}{\partial \alpha} = -\frac{1}{2} \tr( \XiG^{-1} \boK_{,\alpha} ) - \sum_i \frac{f_i}{\alpha f_i +1},
\end{equation}
\begin{equation}
\frac{\partial^2 \ln \p(\bof|\alpha,\beta)}{\partial \alpha^2} = -\frac{1}{2} \tr( \XiG^{-1} \boK_{,\alpha\alpha} ) + \sum_i \frac{f_i^2}{(\alpha f_i +1)^2},
\end{equation}
hence
\begin{equation}
F^f_{\alpha\alpha} = \frac{1}{2} \tr( \XiG^{-1} \left\langle \boK_{,\alpha\alpha} \right\rangle ) - N_\mathrm{pix} \left\langle \boh^2 \right\rangle .
\end{equation}
Using equations \eqref{eq:Kaa} and \eqref{eq:expt_h2}, we obtain
\begin{equation}
F^f_{\alpha \alpha} = \frac{N_\mathrm{pix}}{\alpha^2} \left( 2 - 4\erm^{\alpha^2} + \erm^{3\alpha^2} \right) + \frac{\erm^{2\alpha^2}}{\alpha^4} \tr \left[\XiG^{-1} \exp(\alpha^2 \XiG) \right] + \left[\frac{1}{\alpha^4} (1-2\erm^{\alpha^2}) + 1 + \frac{2}{\alpha^2} (1-\erm^{\alpha^2}) \right] \tr \left( \XiG^{-1} \boldsymbol{1}.\boldsymbol{1}^\intercal \right).
\label{eq:Ff_aa}
\end{equation}
Regarding the cross-term, using the identity $(\XiG^{-1})_{,\beta} = - \XiG^{-1} \XiGbeta \XiG^{-1}$, we get
\begin{equation}
\frac{\partial^2 \ln \p(\bof|\alpha,\beta)}{\partial \alpha \partial \beta} = \frac{1}{2} \tr( \XiG^{-1} \XiGbeta \XiG^{-1} \boK_{,\alpha} ).
\end{equation}
To compute
\begin{equation}
F^f_{\alpha\beta} = -\frac{1}{2} \tr( \XiG^{-1} \XiGbeta \XiG^{-1} \left\langle \boK_{,\alpha} \right\rangle ),
\end{equation}
we use equation \eqref{eq:Ka}, which gives
\begin{equation}
F^f_{\alpha \beta} = \frac{1-\erm^{\alpha^2}}{\alpha} \tr( \XiG^{-1} \XiGbeta ).
\label{eq:Ff_ab}
\end{equation}
The last calculation is similar to the well-known derivation of the Fisher information matrix for Gaussian random fields, appearing in \citet{Tegmark1997}: using $(\ln \XiG)_{,\beta} = \XiG^{-1}\XiGbeta$ and $(\XiG^{-1})_{,\beta} = - \XiG^{-1} \XiGbeta \XiG^{-1}$, 
\begin{gather}
\frac{\partial \ln \p(\bof|\alpha,\beta)}{\partial \beta} = -\frac{1}{2} \tr( \XiG^{-1}\XiGbeta - \XiG^{-1} \XiGbeta \XiG^{-1} \boK ), \\
\frac{\partial^2 \ln \p(\bof|\alpha,\beta)}{\partial \beta^2} = -\frac{1}{2} \tr( -\XiG^{-1} \XiGbeta \XiG^{-1} \XiGbeta + \XiG^{-1}\XiGbetabeta  + 2 \XiG^{-1} \XiGbeta \XiG^{-1} \XiGbeta \XiG^{-1} \boK - \XiG^{-1} \XiGbetabeta \XiG^{-1}\boK ).
\end{gather}
Therefore, using $\left\langle \boK \right\rangle = \XiG$,
\begin{equation}
F^f_\mathrm{\beta\beta} = \frac{1}{2} \tr( \XiG^{-1} \XiGbeta \XiG^{-1} \XiGbeta ).
\label{eq:Ff_bb}
\end{equation}

\subsection{Numerical results}

The Cramér-Rao bounds on $\alpha$ and $\beta$ for a field-level analysis are $\Delta^f \alpha \equiv \sqrt{(F^f)^{-1}_{\alpha\alpha}}$ and $\Delta^f \beta \equiv \sqrt{(F^f)^{-1}_{\beta\beta}}$, where the $(F^f)^{-1}_{ij}$ are the elements of the inverse Fisher matrix, which is defined by equations \eqref{eq:Ff_aa}, \eqref{eq:Ff_ab}, and \eqref{eq:Ff_bb}. Numerically, we find
\begin{align}
& \Delta^f \alpha \approx 0.0099,~\Delta^f \beta \approx 0.0136 \quad \mathrm{for} \quad \alpha =1,~\beta = 0.5, \quad \mathrm{and} \\
& \Delta^f \alpha \approx 0.0352,~\Delta^f \beta \approx 0.0132 \quad \mathrm{for} \quad \alpha =0.2,~\beta = 0.5.
\end{align}

For the fields analysed in this work, the realised marginal errors on $\alpha$ and $\beta$, obtained from the DA samples, are
\begin{align}
& \sigma_\alpha \approx 0.0119,~\sigma_\beta \approx 0.0146 \quad \mathrm{(field~1)}, \\
& \sigma_\alpha \approx 0.0142,~\sigma_\beta \approx 0.0134 \quad \mathrm{(field~2)}, \\
& \sigma_\alpha \approx 0.0119,~\sigma_\beta \approx 0.0138 \quad \mathrm{(field~3)}, \\
& \sigma_\alpha \approx 0.0156,~\sigma_\beta \approx 0.0138 \quad \mathrm{(field~4)}, \\
& \sigma_\alpha \approx 0.0317,~\sigma_\beta \approx 0.0130 \quad \mathrm{(field~5)};
\end{align}
Therefore, with a small but non-zero noise value $\sigma$, we find errors that are typically marginally larger than the predicted bounds, by a few percent. We note that, since our model is non-linear, the Hessian matrix elements scatter around the Fisher matrix elements, so in a given realisation, it is possible to have errors smaller than the Cramér-Rao prediction. This is the case for field 5.\footnote{In another realisation with ground truth parameters $\alpha =0.2$ and $\beta = 0.5$, we obtained $\sigma_\alpha \approx 0.0351$, $\sigma_\beta \approx 0.0140$, in better agreement with the Cramér-Rao bounds.}

\section{Bias in LBA}
\label{apx:Bias in LBA}
\setcounter{figure}{0}
\renewcommand\thefigure{\thesection.\arabic{figure}}

\begin{figure*}
	\begin{center}
    	\includegraphics[width=\textwidth]{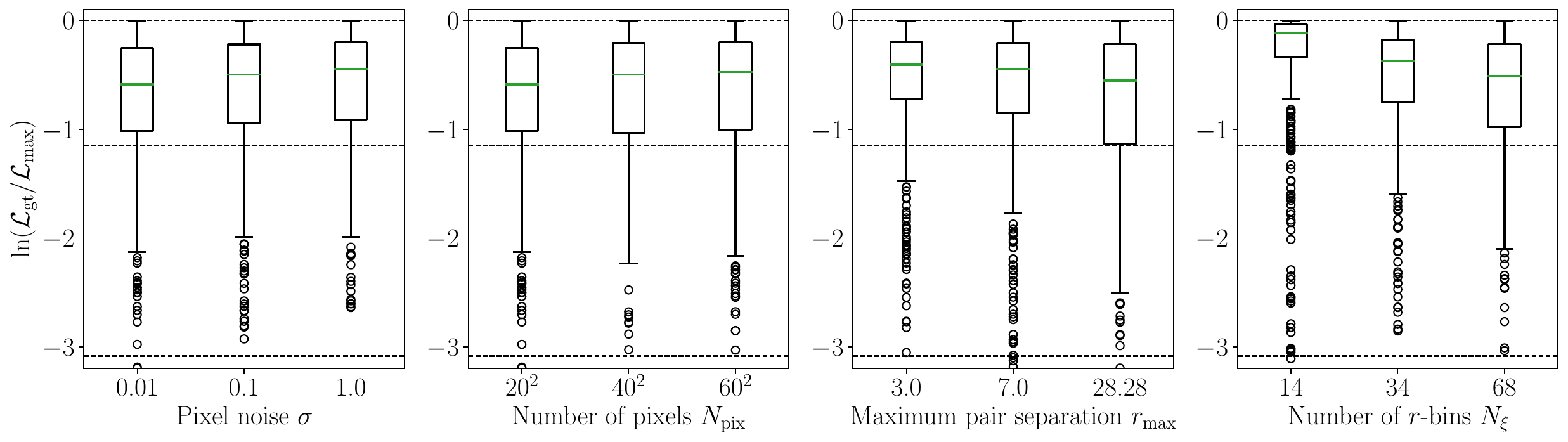}
    \end{center}
    \vspace*{-15pt}
    \caption{
    Box plots showing the distribution of $\ln(\mathcal{L}_\mathrm{gt} / \mathcal{L_\mathrm{max}})$, a proxy for the bias of LBA, as a function of setup parameters. Horizontal dashed lines at $-1.148$ and $-3.090$ correspond to $1$- and $2\sigma$ biases, respectively. Reference values are $\sigma = 0.01$, $N_\mathrm{pix}=20^2$, $r_\mathrm{max}=\sqrt{2N_\mathrm{pix}}$, $N_\xi=68$. In each panel, one parameter, indicated in the $x$-axis, is varied.
    }
    \label{fig:bias_lba}
\end{figure*}

In this appendix, we investigate the impact of some setup parameters on the accuracy of the LBA of the 2PCF of LN fields.

As a measure of accuracy, we use $\ln(\mathcal{L}_\mathrm{gt} / \mathcal{L_\mathrm{max}})$, where $\mathcal{L}_\mathrm{gt} \equiv \p(\boxi^\mathrm{obs}|\alpha,\beta,\boSighat)$ is the likelihood of ground truth parameters, and $\mathcal{L}_\mathrm{max}$ is the maximum likelihood value, which is found by numerical optimisation. The quantity $\ln(\mathcal{L}_\mathrm{gt} / \mathcal{L_\mathrm{max}})$ is easily interpretable: it is non-positive and, using the percent point function of the $\chi^2$ distribution with two degrees of freedom, thresholds at $-1.148$, $-3.090$, $-5.915$ correspond approximately to $1$-, $2$- and $3\sigma$ biases (i.e. the ground truth values out of the $68.3$\%, $95.4$\% and $99.7$\% credible regions, respectively).

Keeping all other parameters fixed at the values used in sections \ref{sec:Model} and \ref{sec:Results}, we investigate the impact of varying the noise $\sigma$, the number of pixels $N_\mathrm{pix}$, the maximum pair separation used in the 2PCF $r_\mathrm{max}$, and the number of $r$-bins $N_\xi$. For each setup, we run $N_\mathrm{sims} = 500$ simulations at ground truth values $\alpha=1$ and $\beta=0.5$ and estimate $\boSighat$ as discussed in section \ref{sec:LBA}. We then examine the distribution of $\ln(\mathcal{L}_\mathrm{gt} / \mathcal{L_\mathrm{max}})$ for each of the $N_\mathrm{sims}$ realisations of $\boxi^\mathrm{obs}$. The results are presented as box plots in figure \ref{fig:bias_lba}. There, the green line shows the median, the box shows the interquartile range $\mathrm{IQR} = \mathrm{Q}_3 - \mathrm{Q}_1$, and the whiskers mark the range of non-outlier data. Outliers (defined as values outside the interval $[\mathrm{Q}_1 - 1.5 \times \mathrm{IQR}, \mathrm{Q}_3 + 1.5 \times \mathrm{IQR}]$) are shown as circles. 

We find no clear dependence of the accuracy of LBA on the noise level $\sigma$ and on the number of pixels $N_\mathrm{pix}$, as can be observed in the two leftmost panels of figure \ref{fig:bias_lba}. On the other hand, we find a dependence on $r_\mathrm{max}$ and $N_\xi$, shown in two rightmost panels of figure \ref{fig:bias_lba}. Using a constant number of $r$-bins $N_\xi$, limiting the analysis of the 2PCF to small scales yields a smaller bias than including all scales up to the maximum pair separation in the image, $r_\mathrm{max} = \sqrt{2N_\mathrm{pix}}$. Similarly, using all scales, using broader $r$-bins (smaller $N_\xi$) leads to less bias. 

From these investigations, we deduce that, by tuning of the 2PCF estimator ($r_\mathrm{max}$ and $N_\xi$), it is possible to Gaussianise the 2PCF likelihood to some degree. Nevertheless, the final accuracy weakly depends on the more fundamental parameters (noise $\sigma$ and image size $N_\mathrm{pix}$). As a consequence, we conclude that our findings regarding the limited accuracy of LBA of the 2PCF will qualitatively hold for real cosmological surveys.

\end{document}